\documentclass{article}
\usepackage{spconf,amsmath,graphicx}
\usepackage{multirow}
\usepackage[table,xcdraw]{xcolor}
\usepackage{url}

\usepackage{algorithm}
\usepackage{algpseudocode}
\usepackage{epsfig}

\title{Darts-Conformer: Towards Efficient Gradient-Based Neural Architecture Search For End-to-End ASR}
\name{Xian Shi$^{1,2*}$\thanks{$^*$Work performed as an intern at Sogou. Lei Xie is the corresponding author.}, Pan Zhou$^2$, Wei Chen$^2$, Lei Xie$^1$}

\address{
  $^1$Audio, Speech and Language Processing Group (ASLP@NPU), \\School of Computer Science,
  Northwestern Polytechnical University, Xi’an, China\\
  $^2$AI Interaction Division, Sogou Inc., Beijing, China}
\begin{document}
%
\maketitle
\begin{abstract}
Neural architecture search (NAS) has been successfully applied to tasks like image classification and language modeling for finding efficient high-performance network architectures. In ASR field especially end-to-end ASR, the related research is still in its infancy. In this work, we focus on applying NAS on the most popular manually designed model: Conformer, and propose an efficient ASR model searching method that benefits from the natural advantage of differentiable architecture search (Darts) in reducing computational overheads. We fuse \emph{Darts mutator} and Conformer blocks to form a complete search space, within which a modified architecture called Darts-Conformer cell is found automatically. The entire searching process on AISHELL-1 dataset costs only 0.7 GPU days. Replacing the Conformer encoder by stacking searched architecture, we get an end-to-end ASR model (named as Darts-Conformner) that outperforms the Conformer baseline by 4.7\% relatively  on the open-source AISHELL-1 dataset. Besides, we verify the transferability of the architecture searched on a small dataset to a larger 2k-hour dataset. 
\end{abstract}
\section{Introduction}

As a main stream of automatic machine learning (AutoML), neural architecture search (NAS) aims at automating the manual process of architecture design~\cite{elsken2019neural}. In recent years, as the state-of-the-art results on tasks like image classification and language modeling have been achieved~\cite{real2017large,zoph2018learning,real2019regularized,baker2016designing}, NAS is attracting attention increasingly. 

Basically, NAS can be classified into three categories: Reinforcement learning (RL) based~\cite{zoph2016neural} searching, evolution algorithm (EA) based~\cite{real2017large} searching and gradient based~\cite{liu2018darts} searching. The former two methods generate a finite number of discrete neural networks with controller RNN or sample a net within a neural network population and train these models from scratch, which raises a relatively high demand for computational overheads 
(2000 GPU days of RL~\cite{zoph2018learning} on CIFAR-10 and 3150 GPU days of evolution~\cite{real2019regularized} on ImageNet). 
On contrast, differentiable architecture search (Darts)~\cite{liu2018darts} is based on the continuous relaxation of the architecture representation, allowing efficient search of the architecture using gradient descent. It eliminates the need to train and compare discrete neural networks. 
Parameters of architecture determination and basic operation
are optimized alternately within a single supernet.

The recent years have seen the rapid development of end-to-end (E2E) models applied in automatic speech recognition (ASR) field,
including connection temporal classification (CTC)~\cite{graves2006connectionist}, recurrent neural network transducer (RNN-T)~\cite{rao2017exploring,li2019improving} and attention based encoder decoder (AED)~\cite{vaswani2017attention,2016Listen,kim2017joint}, etc.
Benefiting from the strong ability of multi-headed-attention (MHA) in global context modeling, transformer~\cite{vaswani2017attention} is competitive among these models in non-streaming tasks, hence kinds of variants come out~\cite{mohamed2019transformers,zhang2020streaming,gulati2020conformer} and become the mainstream of ASR research.

Despite of the excellent performance of AED models, NAS has not been broadly explored in this field for replacing artificial architectures to obtain further improvement, which is the main contribution of our work.
In the work of Conformer~\cite{gulati2020conformer}, the researchers conducted several comparative experiments and ablation experiments around the blocks to find the best architecture. 
However, such process can be done automatically with Darts and a complete search space. 
Inspired by the cell made up of nodes in Darts, we consider the four modules (two feed-forward modules, one convolution module and one MHA module) in Conformer block as nodes, and they can form cell with high degree of freedom. Moreover, the candidate operation within these nodes can be chosen simultaneously. In this paper, we introduce our method to search encoder architecture of end-to-end ASR automatically, which we named as Darts-Conformer. 

The following of the paper is organized as follows. In Section 2, we give a brief review of NAS and Darts algorithms as well as how to combine them with ASR models. Our methods of Darts architecture search for Conformer, including the search space and searching strategies, are presented in Section 3. In Section 4, we discuss the experiments results and analyze the searching process to show the effectiveness of Darts-Conformer. We conclude our work with our findings and future works in Section 5.

\section{Related Work and Background}

\subsection{NAS in speech field}
Kinds of neural architecture search have been applied
in speech field in recent years, especially in speech recognition~\cite{hu2020neural,zheng2020efficient,kim2020evolved,mehrotra2021bench} and keyword spotting~\cite{veniat2019stochastic,mazzawi2019improving,zhang2020autokws}. Evolution algorithm based and gradient based methods are explored in these works. 

Focusing on ASR, there are several valuable explorations. Hu \textit{et al.}~\cite{hu2020neural} used several improved Darts based methods to find efficient context offset and bottleneck dimension for acoustic model. Zheng \textit{et al.}~\cite{zheng2020efficient} proposed straight-through gradient search beyond SNAS~\cite{snas} and ProxylessNAS~\cite{cai2018proxylessnas} and trained acoustic model with CRF-CTC~\cite{xiang2019crf}, but the search space was still limited to receptive field and dimension of TDNN-F only. Kim \textit{et al.}~\cite{kim2020evolved} proposed a relatively complex search space: evolved Transformer, split the encoder and decoder of Transformer into left and right branch and iterated the population with progressive dynamic hurdles (PDH)~\cite{real2019regularized}. The final network is smaller and faster to train, however, the searching progress requires massive computing resources. He \textit{et al.}~\cite{He2020LearnedTA} applied Darts on convolution networks for ASR encoder and yielded considerable improvements over DFSMN baseline. Despite these previous attempts, there still remains space for improvement in search space complexity and search efficiency, and few of these works is applied on novel end-to-end ASR frameworks like Conformer.


\subsection{Preliminary Darts}

Darts~\cite{liu2018darts} proposes a gradient-based architecture searching method for both convolution and recurrent neural networks. For the former case, several normal cells and reduction cells are stacked to build up the final network. A \emph{cell} is a directed acyclic (DAG) graph consisting of an sequence of \emph{nodes}. Each node $x^{(i)}$ is a latent representation and each directed edge $(i,j)$ is associated with some operation $o^{(i,j)}$ that transforms $x^{(i)}$.

In such case we have the output of each node based on all of its predecessors:
\begin{equation}
	x^{(j)} = \sum_{i<j}o^{(i,j)}(x^{i}).
\end{equation}

Darts innovatively relaxes the neural architecture search space and proposes a bi-level optimization method. Let $\mathcal{O}=\{o^{(i,j)}_1,o^{(i,j)}_2,\cdots,o^{(i,j)}_M\}$ be the set of $M$ candidate operations between $x^{(i)}$ and $x^{(j)}$. The output of this edge is mixed with a softmax over all operations:
\begin{equation}
	\overline{o}^{(i,j)}(x)=\sum_{o\in{\mathcal{O}}}\frac{exp(\alpha_o^{(i,j)})}{\sum_{o'\in{\mathcal{O}}}{exp(\alpha_{o'} ^{(i,j)})}}o(x).
\end{equation}

Denote the training loss and validation loss with $\mathcal{L}_{train}$ and $\mathcal{L}_{val}$. The goal of architecture search is to find a set of $\alpha$ that minimizes the $\mathcal{L}_{val}(w^*,\alpha^*)$, where $w^*$ represents the model weights which minimize the training loss $w^*=argmin_w \mathcal{L}_{train}(w,\alpha^*)$. Such optimization problem with both lower-level and upper-level variables implies a bi-level optimization. 
Darts approximates $w^*(\alpha)$ with a single training step of adapting $w$ on current $\alpha$. Thus we have the approximate architecture gradient:
\begin{equation}
\nabla_\alpha\mathcal{L}_{val}(w-\xi\nabla_w\mathcal{L}_{train}(w,\alpha),\alpha),
\label{ag}
\end{equation} 
applying chain rule to Equation~(\ref{ag}) yeilds
\begin{equation}
\nabla_\alpha\mathcal{L}_{val}(w^{'},\alpha)-\xi\nabla^{2}_{\alpha,w}\mathcal{L}_{train}(w,\alpha)\nabla_{w^{'}}\mathcal{L}_{val}(w^{'},\alpha).
\label{ag2}
\end{equation} 
The method in practice is to descend architecture gradient in Equation~(\ref{ag2}) and the model gradient 
$\nabla_w\mathcal{L}_{train}(w,\alpha)$ alternately. The case of $\xi=0$ is referred as \emph{first-order approximation} and the gradient formulation with $\xi>0$ is referred as \emph{second-order approximation}.



\subsection{Conformer}

As aforementioned, many works have been done to strengthen the local context modeling ability of speech Transformer, which is proved to be effective. Conformer~\cite{gulati2020conformer} achieves state-of-the-art results on LibriSpeech, outperforming the previous best published Transformer Transducer by 15\% relative improvement on the test-other dataset. 

The core methods they proposed are replacing the original feed-forward layer in the
Transformer block into two half-step feed-forward layers which is inspired by Macaron-Net~\cite{lu2019understanding}, and using a convolution module which contains a gating mechanism after multi-headed self-attention. 

\section{Methods}
\subsection{Search Space}

A NAS search space is supposed to contain enough reasonable sub-networks. In Darts case, however, it means that the supernet represented by a DAG has a reasonable computing structure and the nodes in it are endowed with enough degree of freedom to mutate. Inspired by Conformer and the CNN cell architecture in Darts, we propose the Darts-Conformer search space, which maintains the backbone components of Conformer encoder and allows gradient-based mutator to adopt NAS within. The core idea of Darts-Conformer cell (denoted as DC-cell) is fusing the operation in Conformer blocks into Darts cell.


\subsubsection{Basic Darts Cell}

\begin{figure}[h]
	\centering
	\includegraphics[width=\linewidth]{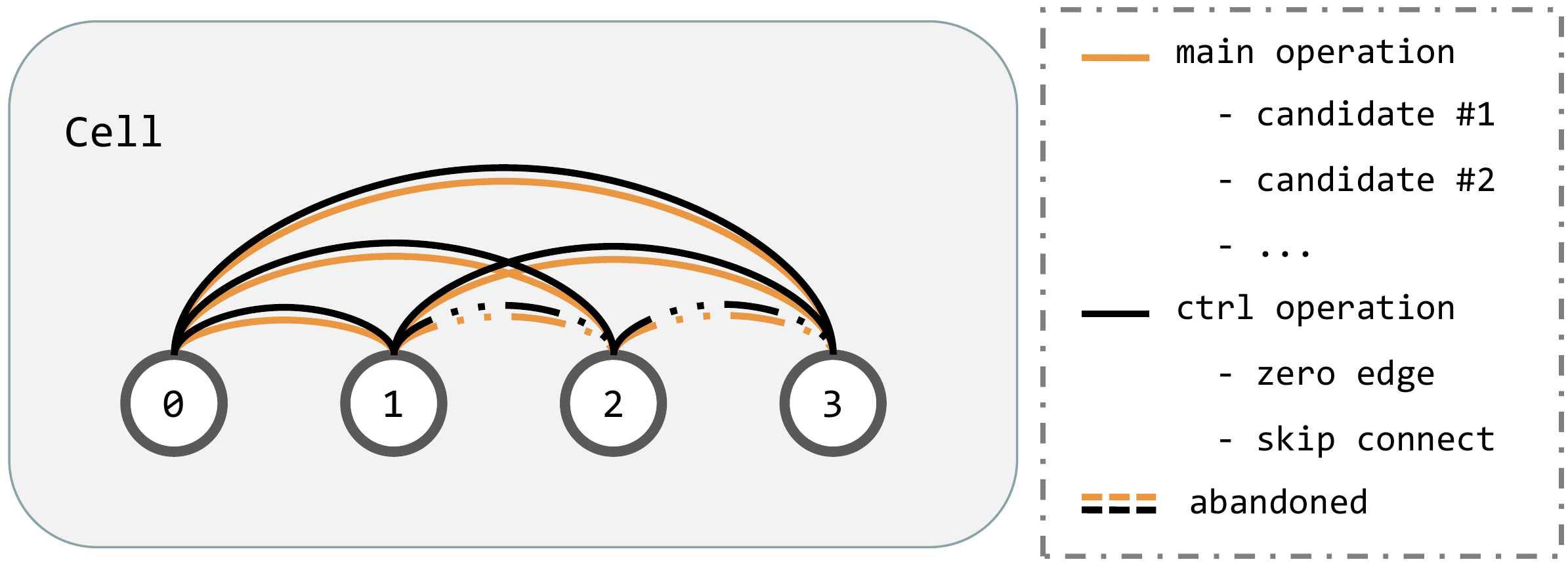}
	\caption{Basic Darts cell architecture.}
	\label{fig:cell}
\end{figure}

Considering a simple cell with four nodes shown in Figure \ref{fig:cell}, each node has all of its predecessor nodes' output as its candidate input. The yellow edge refers to the set of candidate operations while the black edge refers to the combination of zero edge and skip connection edge (for cutting off the path and  generating residual connection respectively). In searching stage, weights of zero edge actually suppress an edge more or less and decide which path to choose. 
Meanwhile, main operations and skip connection competes with each other.

\subsubsection{Darts-Conformer Cell}

\begin{figure}[t]
	\centering
	\includegraphics[width=\linewidth]{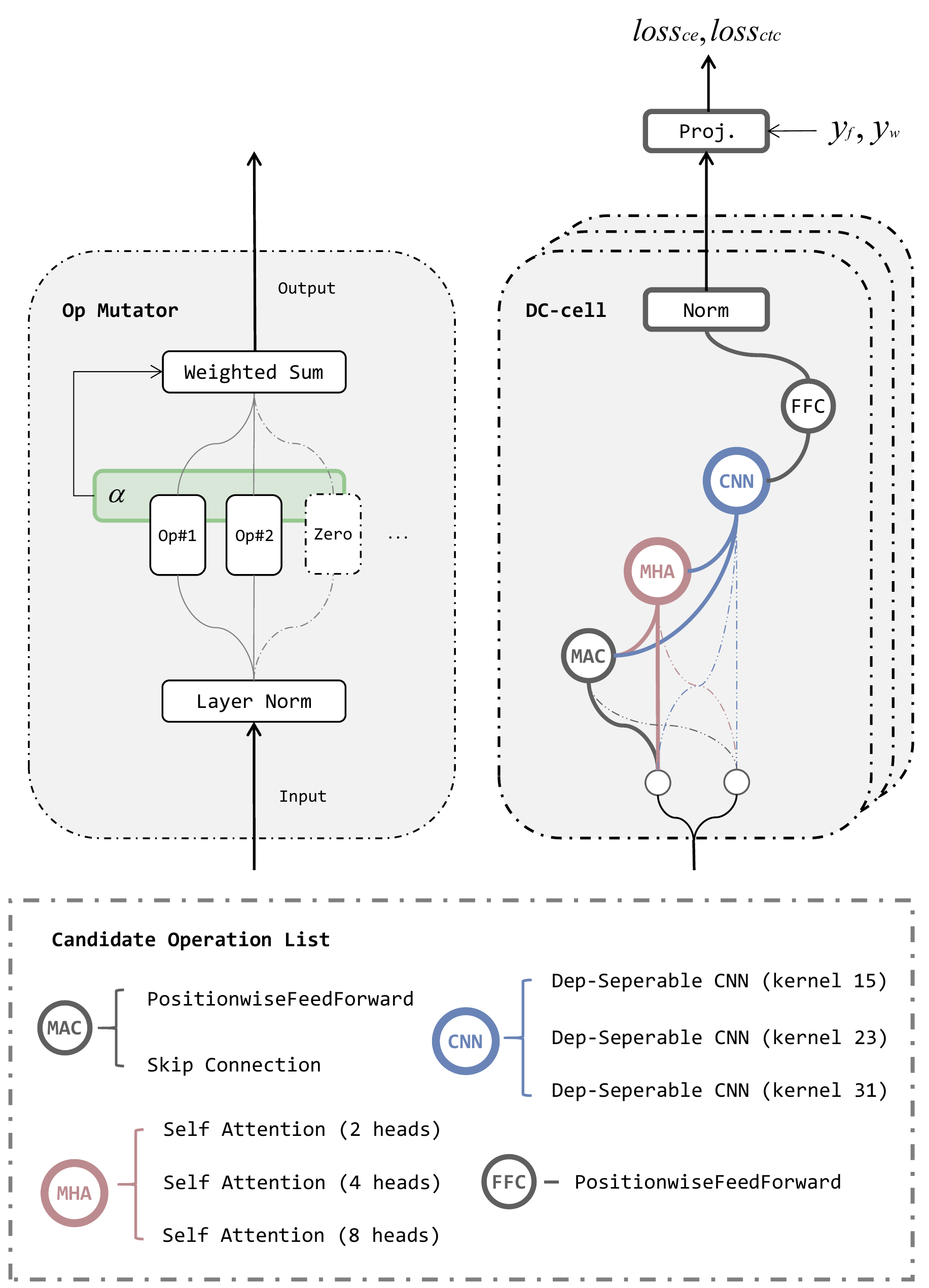}
	\caption{The Darts-Conformer architecture (right). Notice that operations lie on the connection lines of DAG.  Each single line in DC-cell represents an Op mutator (left). The unselected paths (searched on AISHELL-1) are represented by dotted lines.}
	\label{fig:dc}
\end{figure}

Intuitively, Darts-Conformer cell (DC-cell) is constructed by fusing Darts cell into the Conformer-based encoder model which allow the preliminary blocks of Conformer to choose inputs and the containing operation, also to generate residual connections.
As shown in Figure~\ref{fig:dc}, a DC-cell contains four nodes in sequence: Macaron feed forward (MAC), multi-headed attention (MHA), convolution network (CNN) and a feed forward layer (FFC) at last. Their connection order is similar to Conformer, however, the data flow is relatively complex.

\begin{table}[hbp]
	\caption{Input candidates and n\_chosen of Nodes in DC-Cell. Note that both $\rm{Node}_{0}$ and $\rm{Node}_{1}$ represent input nodes, because $\rm{Node}_{MHA}$ may discard the output of $\rm{Node}_{MAC}$.}
	\begin{tabular}{llll}
		\hline
		Index & Name       & Input candidate & n\_chosen \\ \hline
		0,1        & $\rm{Node}_{input}$ & -               & -         \\
		2          & $\rm{Node}_{MAC}$        & \{0,1\}       & 1         \\
		3          & $\rm{Node}_{MHA}$        & \{0,1,2\}     & 2         \\
		4          & $\rm{Node}_{CNN}$        & \{1,2,3\}     & 2         \\
		5          & $\rm{Node}_{FFC}$        & \{4\}       & 1         \\ \hline
	\end{tabular}
	\centering
	\label{cadidate}
\end{table}

Each line in DC-Cell (except the ones towards output) represents of an \emph{operation candidate} which contains a set of operations to choose from and their weights $\alpha$. The input candidates of nodes are listed in Table~\ref{cadidate}. The n\_chosen in the table represents the number of inputs to choose from, and for nodes with multiple inputs, they will sum up the inputs.  For instance, $\rm{Node}_{CNN}$ is the fourth node and it will choose 2 from outputs of $\rm{Node}_{MAC}$, $\rm{Node}_{MHA}$ and $\rm{Node}_{input}$.

As for candidate operations, we focus on the kernel size of CNN block (choose from \{15, 23, 31\}) and the number of heads in the self-attention. Besides, we try to find out that if macaron feed forward will be kept by Darts.

\subsection{Searching Strategy}

Different from applying convolution networks on image classification tasks, there is a big variation of the speech data over $T$-dimension and E2E ASR models usually contain massive parameters, which tends to exacerbates the \emph{collapse issue} in the early stage of searching~\cite{liang2019darts+}. Denoting the architecture-determining parameters as $\alpha$ and parameters of basic operation as $w$, before optimizing $\alpha$ and $w$ alternatively within a single step, we freeze $\alpha$ for $e_{w}$ epochs to ensure the operations to pass the initial convergence stage fairly (as shown in Algorithm \ref{alg:routine}). In practice, we use first-order approximation and adam optimizer in searching.

\begin{algorithm}[htb] 
\caption{ Searching and training routine of Dart-Conformer.} 
\label{alg:routine} 
\begin{algorithmic}[1] 
\Require 
\State Form the supernet $S$ with the candidate operation and alternative inputs.
\State Freeze the mutator parameters $\alpha$ for $e_{w}$ epochs and update $w$ with CTC loss.
\State Use first-order approximation to update $\alpha$ and $w$ alternatively with CTC loss.
\\\Return the found mutator parameters $\alpha^*$.
\Ensure 
\State Fix the architecture of a single cell $S^{'}$ with $\alpha^*$ (abort the unselected operations and connections from $S$).
\State Stack the cell into multiple-layer encoder to compose an E2E ASR model $M$; \\
\Return trained $M$ as the final model and decode; 
\end{algorithmic} 
\end{algorithm}

\subsection{Strengths of Our Work}
Comparing to the previous studies about NAS in ASR aforementioned, our method has advantages in the following aspects:
\begin{itemize}
  \item [1)] 
  The proposed search space absorbs both advantages of Darts cell in architecture selection and the strengths of Conformer in local and global context modeling.
  \item [2)]
  The search process of gradient based methods is significantly shorter than those of RL and evolution based methods.
  \item [3)]
  We conduct Darts search on encoder architecture with CTC loss and frame-level CE loss rather than the entire encoder-decoder model which eliminates the potential impact from the source attention and the decoder.
\end{itemize}

\section{Experiments}

We follow the route below for NAS experiments on ASR task: (i) search on a small dataset; (ii) stack a single searched architecture to make up an encoder; (iii) validate the stacked encoder on the same dataset; (iv) evaluate the transferability of the stacked encoder on an unseen and larger dataset.

\subsection{Dataset}

We search on the open AISHELL-1 data \cite{bu2017aishell} to find the best DC-cell architecture. This data contains 176 hours of Mandarin speech collected from 400 speakers, and 10 hours develop set.
The selected Darts-Conformer is retrained on the same data to exploit the ASR performance of the selected DC-cell. A bigger dataset that contains 2,000 hours of Mandarin speech collected from multi-domain by Sogou (denoted as Sogou-2k) is used to verify the transferability of Darts-Conformer. We split 30 hours of speech data from Sogou-2k as development set for model training. Three types of test set are used for recognition performance comparison, which consist of 18,822 clean read speech utterances, 11,249 noisy read utterances and 20,000 utterances of far field spontaneous speech. We denote them as Clean, Noisy, and Far-field respectively.


Eighty-dimension of FBank features with 3-dimension pitch features are extracted for AISHELL-1 task and 71-dimension FBank features are extracted for Sogou-2K. During ASR training stage of AISHELL-1, a 3-fold speed perturbation ($\times{0.9}$, $\times{1.0}$, $\times{1.1}$) is applied.

\subsection{Architecture Searching}

We implement the supernet based on the NNI toolkit~\cite{NNI}. In the searching stage, we use feed forward layers with 512-dimension hidden units. We use two separate adam optimizers for $\alpha$ and $\theta$, learning rate is $0.0002$ and $0.0003$ respectively. We use the mixture of CTC loss (0.7) and frame-level CE loss (0.3) for gradient descent in Darts. And weights $\alpha$ is frozen for 3 epochs. The maximum epoch is 30 and there are 1.6k steps in a epoch under the batch size of 48. An entire searching experiment under the setup above with a single Titan Xp GPU costs 16 hours. 

\begin{figure}[h]
	\centering
	\includegraphics[width=\linewidth]{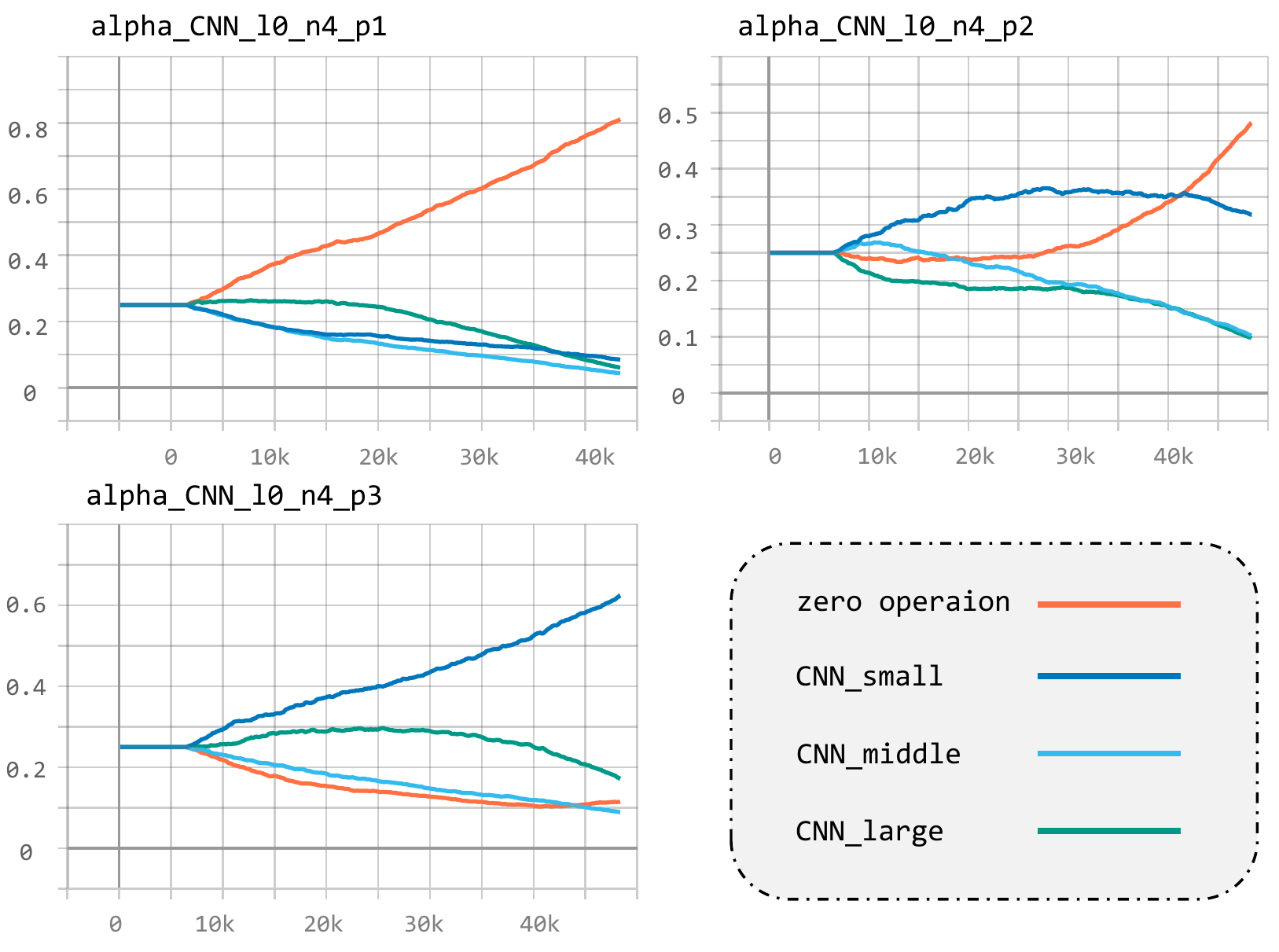}
	\caption{Operation weights $\alpha$ of $\rm{Node}_{CNN}$ while searching. p1, p2, p3 are corresponding to input candidate [1,2,3] in Table~\ref{cadidate}. }
	\label{fig:cnn}
\end{figure}

Figure~\ref{fig:cnn} shows how weights $\alpha$ in $\rm{Node}_{CNN}$ change in the training process. The curves of zero operation (orange) indicate that the input from $\rm{Node}_{input}$ is inhibited mostly. Thus a extra CNN operation using the raw input is added to preliminary Conformer. And it's clear that Darts tends to choose CNN with smaller kernels for Conformer, which is consistent with the previous results of some artificial network. As for $\rm{Node}_{MHA}$, the recommended number of heads in self-attention is 4 for both. In our series of experiments, we find that self-attention blocks are abandoned early in Darts search, which means that the input candidate determined by zero operation in $\rm{Node}_{MHA}$ might be stochastic, but the operation candidate mutator works well. Under such circumstance, Darts chooses the outputs from $\rm{Node}_{input}$ and $\rm{Node}_{CNN}$ as input, which means an extra MHA operation to the input without fed into the convolution layer (comparing to Conformer).

\begin{figure}[h]
	\centering
	\includegraphics[width=7.5cm]{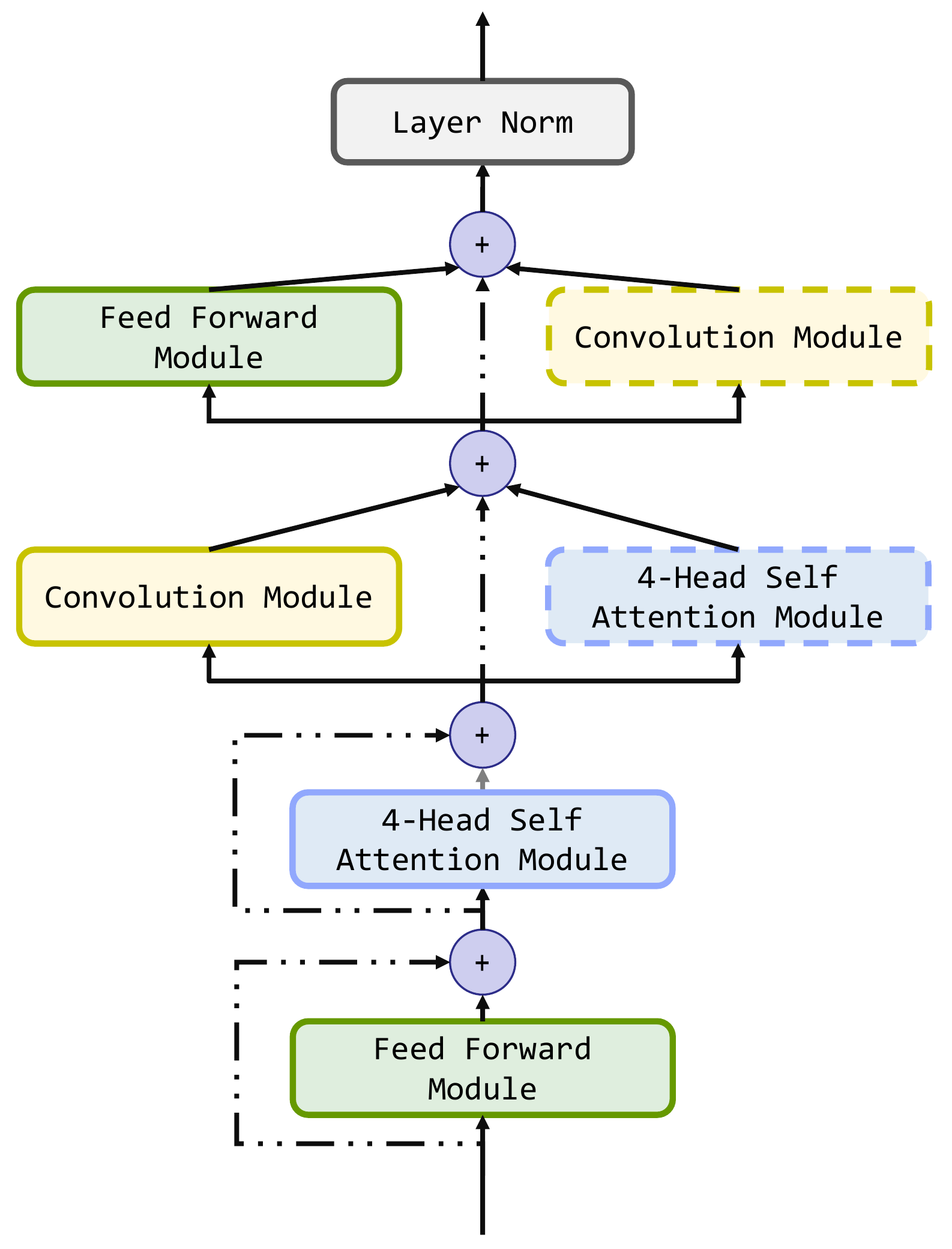}
	\caption{Seached Darts-Conformer model architecture. The kernel size of convolution modules is both 15 and the number of attention modules head is both 4. Dotted boxes refer to adding operations searched from Darts comparing to preliminary Conformer.}
	\label{fig:searched}
\end{figure}

\subsection{ASR Model Training Setup}

We conduct baseline Conformer experiments with ESPnet toolkit~\cite{watanabe2018espnet}. The baseline model contains a 6-layer Conformer encoder and a 3-layer self-attention decoder, and the attention in the model is of 256-dimension with 4 heads. The kernel size of the convolution layer is 31. The encoder-decoder model is trained jointly with CTC loss (joint-weight=0.3) for 65 epochs using adam optimizer and Noam learning rate strategy~\cite{vaswani2017attention}. In the inference stage, we set CTC weight to 0.3 with beam size of 5. No language model is  used in decoding.

For evaluating the performance of Darts-Conformer, we firstly fix both the chosen edges in the DAG and the operations in the mutator, and then stack the single cell 6 times to make up a 6-layer encoder. The training strategies and setups are exactly same as the baseline experiments.
\subsection{Results on AISHELL-1}

We then build up  Darts-Conformer model by stacking several DC-cells in encoder and several self-attention layers in decoder.
The results of the baseline model and proposed Darts-Conformer model in character error rate (CER) is reported in Tabel~\ref{aishell}. The stacked Darts-Conformer cells perform better than the baseline Conformer 7.0\% and 4.7\% respectively on the development set and test set, which proves the effectiveness of the input candidate and operation candidate mutator of Darts. Considering that two edges selected by darts introduce additional parameters, we 
increase the attention dimension in the baseline experiment from 256 to 288 as an comparative experiment and the parameters of the model come up to 27.67M. Under such circumstance, the performance of Darts-Conformer is similar to baseline Conformer with less parameters.

\begin{table}[h]
\caption{Results on AISHELL-1.}
\centering
\label{aishell}
\begin{tabular}{llll}
\hline
& \multicolumn{1}{c}{\multirow{2}{*}{\#para.}} & \multicolumn{2}{c}{CER}         \\ \cline{3-4} 
& \multicolumn{1}{c}{}                         & eval.          & test           \\ \hline
Conformer Baseline                 & 24.17M                                       & 5.7\%          & 6.4\%          \\
~~~~+ larger adim                      & 27.67M                                       & 5.3\%          & 6.1\%          \\ \hline
Darts-Conformer (proposed) & 26.81M                                       &         5.3\% &          6.1\% \\ 
~~~~- input candidate (ablation)       & 26.81M                                       & 5.4\%          & 6.2\%          \\
~~~~- chosen operation (ablation)     & 26.86M                                       & 5.3\%          & 6.3\%          \\ \hline
\end{tabular}
\end{table}

In addition, two ablation experiments are conducted. If the additional edges do not take the output of precursor nodes as Darts prefers and goes in the original order (parallel to the origin layers), the performance of the model will decrease accordingly. On the other hand, if we do not use the chosen operations (small CNN kernels and attention heads of 4), the CERs on both sets raise too.

\subsection{Transferability On Sogou-2k}

For evaluating both the transferability of the architecture searched from AISHELL-1 and the performance on larger dataset, we compare the baseline model and Darts-Conformer on Sogou-2k. The training strategies are same as we use in AISHELL-1 experiments except maximum epoch is set to 30. We evaluate the models on test sets and the results in shown in Table~\ref{2k}. It turns out that Darts-Conformer outperforms the baseline Conformer model in the Clean and Noisy test sets. But in the Far-field test set, they have the same performance.
\begin{table}[h]
\centering
\caption{Transferability experiment results on Sogou-2k.}
\begin{tabular}{lcccc}
\hline
                   & \multicolumn{3}{c}{test sets}          \\ \cline{2-4} 
                   & Clean   & Noisy  &  Far-field   \\ \hline
Conformer Baseline & 10.9\% & 7.5\%  & 14.0\% \\
Darts-Conformer    & 10.6\% & 7.2\%  & 14.0\% \\ \hline
\end{tabular}
\label{2k}
\end{table}

\section{Conclusion}

In this paper we propose Darts-Conformer, a gradient based neural architecture search framework within which we can find better ASR model architectures automatically. We fuse Darts mutator and Conformer efficiently: maintaining the local and global modeling abilities of Conformer and making the nodes able to be connected with each other in a high degree of freedom. Applying Darts in the proposed search space, the searching process costs only 0.7 GPU day and finds the preferring architecture for ASR.
A series of experiments on both AISHELL-1 and Sogou-2k prove that the automatically found architecture outperforms the preliminary Conformer which is a strong baseline. We verify the effectiveness of gradient-based architecture search in ASR field.
We are going deep into the gradient-based search methods in E2E ASR. And we also attempt to design an elaborate and unified supernet where the decoder architecture can also be searched at the same time.

\section{Acknowledgements}
The authors gratefully acknowledge the ﬁnancial support provided by the National Key Research and Development Program of China (2020AAA0108004).

\bibliographystyle{IEEEbib}
\bibliography{strings,refs}

\begin{thebibliography}{10}

\bibitem{elsken2019neural}
Thomas Elsken, Jan~Hendrik Metzen, Frank Hutter, et~al.,
\newblock ``Neural architecture search: A survey.,''
\newblock {\em J. Mach. Learn. Res.}, vol. 20, no. 55, pp. 1--21, 2019.

\bibitem{real2017large}
Esteban Real, Sherry Moore, Andrew Selle, Saurabh Saxena, Yutaka~Leon Suematsu,
  Jie Tan, Quoc~V Le, and Alexey Kurakin,
\newblock ``Large-scale evolution of image classifiers,''
\newblock in {\em International Conference on Machine Learning}. PMLR, 2017,
  pp. 2902--2911.

\bibitem{zoph2018learning}
Barret Zoph, Vijay Vasudevan, Jonathon Shlens, and Quoc~V Le,
\newblock ``Learning transferable architectures for scalable image
  recognition,''
\newblock in {\em Proceedings of the IEEE conference on computer vision and
  pattern recognition}, 2018, pp. 8697--8710.

\bibitem{real2019regularized}
Esteban Real, Alok Aggarwal, Yanping Huang, and Quoc~V Le,
\newblock ``Regularized evolution for image classifier architecture search,''
\newblock in {\em Proceedings of the aaai conference on artificial
  intelligence}, 2019, vol.~33, pp. 4780--4789.

\bibitem{baker2016designing}
Bowen Baker, Otkrist Gupta, Nikhil Naik, and Ramesh Raskar,
\newblock ``Designing neural network architectures using reinforcement
  learning,''
\newblock {\em arXiv preprint arXiv:1611.02167}, 2016.

\bibitem{zoph2016neural}
Barret Zoph and Quoc~V Le,
\newblock ``Neural architecture search with reinforcement learning,''
\newblock {\em arXiv preprint arXiv:1611.01578}, 2016.

\bibitem{liu2018darts}
Hanxiao Liu, Karen Simonyan, and Yiming Yang,
\newblock ``Darts: Differentiable architecture search,''
\newblock {\em arXiv preprint arXiv:1806.09055}, 2018.

\bibitem{graves2006connectionist}
Alex Graves, Santiago Fern{\'a}ndez, Faustino Gomez, and J{\"u}rgen
  Schmidhuber,
\newblock ``Connectionist temporal classification: labelling unsegmented
  sequence data with recurrent neural networks,''
\newblock in {\em Proceedings of the 23rd international conference on Machine
  learning}, 2006, pp. 369--376.

\bibitem{rao2017exploring}
Kanishka Rao, Ha{\c{s}}im Sak, and Rohit Prabhavalkar,
\newblock ``Exploring architectures, data and units for streaming end-to-end
  speech recognition with rnn-transducer,''
\newblock in {\em 2017 IEEE Automatic Speech Recognition and Understanding
  Workshop (ASRU)}. IEEE, 2017, pp. 193--199.

\bibitem{li2019improving}
Jinyu Li, Rui Zhao, Hu~Hu, and Yifan Gong,
\newblock ``Improving rnn transducer modeling for end-to-end speech
  recognition,''
\newblock in {\em 2019 IEEE Automatic Speech Recognition and Understanding
  Workshop (ASRU)}. IEEE, 2019, pp. 114--121.

\bibitem{vaswani2017attention}
Ashish Vaswani, Noam Shazeer, Niki Parmar, Jakob Uszkoreit, Llion Jones,
  Aidan~N Gomez, Lukasz Kaiser, and Illia Polosukhin,
\newblock ``Attention is all you need,''
\newblock {\em arXiv preprint arXiv:1706.03762}, 2017.

\bibitem{2016Listen}
W.~Chan, N.~Jaitly, Q.~Le, and O.~Vinyals,
\newblock ``Listen, attend and spell: A neural network for large vocabulary
  conversational speech recognition,''
\newblock in {\em 2016 IEEE International Conference on Acoustics, Speech and
  Signal Processing (ICASSP)}, 2016.

\bibitem{kim2017joint}
Suyoun Kim, Takaaki Hori, and Shinji Watanabe,
\newblock ``Joint ctc-attention based end-to-end speech recognition using
  multi-task learning,''
\newblock in {\em 2017 IEEE international conference on acoustics, speech and
  signal processing (ICASSP)}. IEEE, 2017, pp. 4835--4839.

\bibitem{mohamed2019transformers}
Abdelrahman Mohamed, Dmytro Okhonko, and Luke Zettlemoyer,
\newblock ``Transformers with convolutional context for asr,''
\newblock {\em arXiv preprint arXiv:1904.11660}, 2019.

\bibitem{zhang2020streaming}
Shiliang Zhang, Zhifu Gao, Haoneng Luo, Ming Lei, Jie Gao, Zhijie Yan, and Lei
  Xie,
\newblock ``Streaming chunk-aware multihead attention for online end-to-end
  speech recognition,''
\newblock {\em arXiv preprint arXiv:2006.01712}, 2020.

\bibitem{gulati2020conformer}
Anmol Gulati, James Qin, Chung-Cheng Chiu, Niki Parmar, Yu~Zhang, Jiahui Yu,
  Wei Han, Shibo Wang, Zhengdong Zhang, Yonghui Wu, et~al.,
\newblock ``Conformer: Convolution-augmented transformer for speech
  recognition,''
\newblock {\em arXiv preprint arXiv:2005.08100}, 2020.

\bibitem{hu2020neural}
Shoukang Hu, Xurong Xie, Shansong Liu, Mengzhe Geng, Xunying Liu, and Helen
  Meng,
\newblock ``Neural architecture search for speech recognition,''
\newblock {\em arXiv preprint arXiv:2007.08818}, 2020.

\bibitem{zheng2020efficient}
Huahuan Zheng, Keyu An, and Zhijian Ou,
\newblock ``Efficient neural architecture search for end-to-end speech
  recognition via straight-through gradients,''
\newblock {\em arXiv preprint arXiv:2011.05649}, 2020.

\bibitem{kim2020evolved}
Jihwan Kim, Jisung Wang, Sangki Kim, and Yeha Lee,
\newblock ``Evolved speech-transformer: Applying neural architecture search to
  end-to-end automatic speech recognition,''
\newblock {\em Proc. Interspeech 2020}, pp. 1788--1792, 2020.

\bibitem{mehrotra2021bench}
Abhinav Mehrotra, Alberto~Gil Ramos, Sourav Bhattacharya, {\L}ukasz Dudziak,
  Ravichander Vipperla, Thomas Chau, Mohamed~S Abdelfattah, Samin Ishtiaq, and
  Nicholas~D Lane,
\newblock ``Nas-bench-asr: Reproducible neural architecture search for speech
  recognition,''
\newblock in {\em International Conference on Learning Representations (ICLR)},
  2021.

\bibitem{veniat2019stochastic}
Tom V{\'e}niat, Olivier Schwander, and Ludovic Denoyer,
\newblock ``Stochastic adaptive neural architecture search for keyword
  spotting,''
\newblock in {\em ICASSP 2019-2019 IEEE International Conference on Acoustics,
  Speech and Signal Processing (ICASSP)}. IEEE, 2019, pp. 2842--2846.

\bibitem{mazzawi2019improving}
Hanna Mazzawi, Xavi Gonzalvo, Aleks Kracun, Prashant Sridhar, Niranjan
  Subrahmanya, Ignacio Lopez-Moreno, Hyun-Jin Park, and Patrick Violette,
\newblock ``Improving keyword spotting and language identification via neural
  architecture search at scale.,''
\newblock in {\em INTERSPEECH}, 2019, pp. 1278--1282.

\bibitem{zhang2020autokws}
Bo~Zhang, WenFeng Li, Qingyuan Li, Weiji Zhuang, Xiangxiang Chu, and Yujun
  Wang,
\newblock ``Autokws: Keyword spotting with differentiable architecture
  search,''
\newblock {\em arXiv preprint arXiv:2009.03658}, 2020.

\bibitem{snas}
Sirui Xie, Hehui Zheng, Chunxiao Liu, and Liang Lin,
\newblock ``Snas: stochastic neural architecture search,''
\newblock {\em arXiv preprint arXiv:1812.09926}, 2018.

\bibitem{cai2018proxylessnas}
Han Cai, Ligeng Zhu, and Song Han,
\newblock ``Proxylessnas: Direct neural architecture search on target task and
  hardware,''
\newblock {\em arXiv preprint arXiv:1812.00332}, 2018.

\bibitem{xiang2019crf}
Hongyu Xiang and Zhijian Ou,
\newblock ``Crf-based single-stage acoustic modeling with ctc topology,''
\newblock in {\em ICASSP 2019-2019 IEEE International Conference on Acoustics,
  Speech and Signal Processing (ICASSP)}. IEEE, 2019, pp. 5676--5680.

\bibitem{He2020LearnedTA}
Liqiang He, Dan Su, and Dong Yu,
\newblock ``Learned transferable architectures can surpass hand-designed
  architectures for large scale speech recognition,''
\newblock {\em ArXiv}, vol. abs/2008.11589, 2020.

\bibitem{lu2019understanding}
Yiping Lu, Zhuohan Li, Di~He, Zhiqing Sun, Bin Dong, Tao Qin, Liwei Wang, and
  Tie-Yan Liu,
\newblock ``Understanding and improving transformer from a multi-particle
  dynamic system point of view,''
\newblock {\em arXiv preprint arXiv:1906.02762}, 2019.

\bibitem{liang2019darts+}
Hanwen Liang, Shifeng Zhang, Jiacheng Sun, Xingqiu He, Weiran Huang, Kechen
  Zhuang, and Zhenguo Li,
\newblock ``Darts+: Improved differentiable architecture search with early
  stopping,''
\newblock {\em arXiv preprint arXiv:1909.06035}, 2019.

\bibitem{bu2017aishell}
Hui Bu, Jiayu Du, Xingyu Na, Bengu Wu, and Hao Zheng,
\newblock ``Aishell-1: An open-source mandarin speech corpus and a speech
  recognition baseline,''
\newblock in {\em 2017 20th Conference of the Oriental Chapter of the
  International Coordinating Committee on Speech Databases and Speech I/O
  Systems and Assessment (O-COCOSDA)}. IEEE, 2017, pp. 1--5.

\bibitem{NNI}
Microsoft,
\newblock ``Neural network intelligence ({NNI}),''
  \url{http://nni.readthedocs.io}.

\bibitem{watanabe2018espnet}
Shinji Watanabe, Takaaki Hori, Shigeki Karita, Tomoki Hayashi, Jiro Nishitoba,
  Yuya Unno, Nelson {Enrique Yalta Soplin}, Jahn Heymann, Matthew Wiesner,
  Nanxin Chen, Adithya Renduchintala, and Tsubasa Ochiai,
\newblock ``{ESPnet}: End-to-end speech processing toolkit,''
\newblock in {\em Proceedings of Interspeech}, 2018, pp. 2207--2211.

\end{thebibliography}

\end{document}